\documentclass[aps,prl,twocolumn,superscriptaddress,showpacs,preprintnumbers,amsmath,amssymb,showkeys,nofootinbib]{revtex4-1}
\usepackage{graphicx}
\usepackage{dcolumn}
\usepackage{bm}
\usepackage{notes2bib}
\usepackage{graphicx}
\usepackage{soul}
\usepackage{gensymb}
\usepackage{url}
\usepackage{lineno}

\usepackage{natbib,hyperref}  
\usepackage{amsmath}
\usepackage{latexsym}
\usepackage{amssymb}
\usepackage{color}
\usepackage[normalem]{ulem}

\usepackage{slashed}

\newcommand*{\ANL}{Argonne National Laboratory, Argonne, Illinois 60439}
\newcommand*{\CSUDH}{California State University, Dominguez Hills, Carson, CA 90747}
\newcommand*{\CANISIUS}{Canisius College, Buffalo, NY}
\newcommand*{\SACLAY}{IRFU, CEA, Universit\'e Paris-Saclay, F-91191 Gif-sur-Yvette, France}
\newcommand*{\CNU}{Christopher Newport University, Newport News, Virginia 23606}
\newcommand*{\UCONN}{University of Connecticut, Storrs, Connecticut 06269}
\newcommand*{\DUKE}{Duke University, Durham, North Carolina 27708-0305}
\newcommand*{\DUQUESNE}{Duquesne University, 600 Forbes Avenue, Pittsburgh, PA 15282 }
\newcommand*{\FU}{Fairfield University, Fairfield CT 06824}
\newcommand*{\FERRARAU}{Universit\`a di Ferrara , 44121 Ferrara, Italy}
\newcommand*{\FIU}{Florida International University, Miami, Florida 33199}
\newcommand*{\GWUI}{The George Washington University, Washington, DC 20052}
\newcommand*{\GSIFFN}{GSI Helmholtzzentrum fur Schwerionenforschung GmbH, D-64291 Darmstadt, Germany}
\newcommand*{\INFNFE}{INFN, Sezione di Ferrara, 44100 Ferrara, Italy}
\newcommand*{\INFNFR}{INFN, Laboratori Nazionali di Frascati, 00044 Frascati, Italy}
\newcommand*{\INFNGE}{INFN, Sezione di Genova, 16146 Genova, Italy}
\newcommand*{\INFNRO}{INFN, Sezione di Roma Tor Vergata, 00133 Rome, Italy}
\newcommand*{\INFNTUR}{INFN, Sezione di Torino, 10125 Torino, Italy}
\newcommand*{\INFNCAT}{INFN, Sezione di Catania, 95123 Catania, Italy}
\newcommand*{\INFNPAV}{INFN, Sezione di Pavia, 27100 Pavia, Italy}
\newcommand*{\ORSAY}{Universit\'{e} Paris-Saclay, CNRS/IN2P3, IJCLab, 91405 Orsay, France}
\newcommand*{\Juelich}{Institute fur Kernphysik (Juelich), Juelich, Germany}
\newcommand*{\JMU}{James Madison University, Harrisonburg, Virginia 22807}
\newcommand*{\KNU}{Kyungpook National University, Daegu 41566, Republic of Korea}
\newcommand*{\LAMAR}{Lamar University, 4400 MLK Blvd, PO Box 10046, Beaumont, Texas 77710}
\newcommand*{\MIT}{Massachusetts Institute of Technology, Cambridge, Massachusetts  02139-4307}
\newcommand*{\MISS}{Mississippi State University, Mississippi State, MS 39762-5167}
\newcommand*{\ITEP}{National Research Centre Kurchatov Institute - ITEP, Moscow, 117259, Russia}
\newcommand*{\UNH}{University of New Hampshire, Durham, New Hampshire 03824-3568}
\newcommand*{\NMSU}{New Mexico State University, PO Box 30001, Las Cruces, NM 88003, USA}
\newcommand*{\NSU}{Norfolk State University, Norfolk, Virginia 23504}
\newcommand*{\OHIOU}{Ohio University, Athens, Ohio  45701}
\newcommand*{\ODU}{Old Dominion University, Norfolk, Virginia 23529}
\newcommand*{\JLUGiessen}{II. Physikalisches Institut der Universit\"at
Gie{\ss}en, 35392 Gie{\ss}en, Germany}
\newcommand*{\ROMAII}{Universit\`a di Roma Tor Vergata, 00133 Rome Italy}
\newcommand*{\MSU}{Skobeltsyn Institute of Nuclear Physics, Lomonosov Moscow State University, 119234 Moscow, Russia}
\newcommand*{\SCAROLINA}{University of South Carolina, Columbia, South Carolina 29208}
\newcommand*{\TEMPLE}{Temple University,  Philadelphia, PA 19122 }
\newcommand*{\JLAB}{Thomas Jefferson National Accelerator Facility, Newport News, Virginia 23606}
\newcommand*{\UTFSM}{Universidad T\'{e}cnica Federico Santa Mar\'{i}a, Casilla 110-V Valpara\'{i}so, Chile}
\newcommand*{\BRESCIA}{Universit\`{a} degli Studi di Brescia, 25123 Brescia, Italy}
\newcommand*{\MESSU}{Universit\`{a} degli Studi di Messina, 98166 Messina, Italy}
\newcommand*{\UCR}{University of California Riverside, 900 University Avenue, Riverside, CA 92521, USA}
\newcommand*{\GLASGOW}{University of Glasgow, Glasgow G12 8QQ, United Kingdom}
\newcommand*{\YORK}{University of York, York YO10 5DD, United Kingdom}
\newcommand*{\VIRGINIA}{University of Virginia, Charlottesville, Virginia 22901}
\newcommand*{\WM}{College of William and Mary, Williamsburg, Virginia 23187-8795}
\newcommand*{\YEREVAN}{Yerevan Physics Institute, 375036 Yerevan, Armenia}
\newcommand*{\ISU}{Idaho State University, Pocatello, Idaho 83209}
{}

\begin{document}
\title{First observation of correlations between spin and transverse momenta in back-to-back dihadron production at CLAS12}

\author{H.~Avakian}
\affiliation{\JLAB}
\author{T.B.~Hayward}
\affiliation{\UCONN}
\author {A. Kotzinian} 
\affiliation{\YEREVAN}
\affiliation{\INFNTUR}
\author {W.R.~Armstrong} 
\affiliation{\ANL}
\author {H.~Atac} 
\affiliation{\TEMPLE}
\author {C.~Ayerbe~Gayoso} 
\affiliation{\WM}
\author {L.~Baashen} 
\affiliation{\FIU}
\author {N.A.~Baltzell} 
\affiliation{\JLAB}
\author {L. Barion} 
\affiliation{\INFNFE}
\author {M. Bashkanov} 
\affiliation{\YORK}
% ...
\collaboration{The CLAS Collaboration}

\date{\today}
\begin{abstract}
We report the first measurements of deep inelastic scattering spin-dependent azimuthal asymmetries in back-to-back dihadron electroproduction, where two hadrons are produced in opposite hemispheres along the z-axis in the center-of-mass frame, with the first hadron produced in the current-fragmentation region and the second in the target-fragmentation region. The data were taken with longitudinally polarized electron beams of 10.2 and 10.6~GeV incident on an unpolarized liquid-hydrogen target using the CLAS12 spectrometer at Jefferson Lab. Observed non-zero $\sin\Delta\phi$ modulations in $ep \rightarrow e'p\pi^+X$ events, where $\Delta\phi$ is the difference of the azimuthal angles of the proton and pion in the virtual photon and target nucleon center-of-mass frame, indicate that correlations between the spin and transverse momenta of hadrons produced in the target- and current-fragmentation regions may be significant. The measured beam-spin asymmetries provide a first access in dihadron production to a previously unobserved leading-twist spin- and transverse-momentum-dependent fracture function. The fracture functions describe the hadronization of the target remnant after the hard scattering of a virtual photon off a quark in the target particle and provide a new avenue for studying nucleonic structure and hadronization.
\end{abstract} 
\setcounter{footnote}{0}
\maketitle

\section{Introduction}
The quest for a complete understanding of nucleonic structure and the mechanism by which hadrons form out of constituent partons is one of the ultimate goals of nuclear physics. In deep inelastic scattering (DIS) an electron scatters off a nucleon with sufficient energy and momentum transfer that the process is well-described by incoherent scattering from individual partons (quarks or gluons). In semi-inclusive deep inelastic scattering (SIDIS), one or more hadrons are detected in coincidence with the scattered electron, providing information on the initial quark flavor, transverse momentum and spin~\cite{Anselmino:2020vlp}. The majority of SIDIS studies have focused on the analysis of hadron production in the current-fragmentation region (CFR), where the final-state hadrons are produced from the struck quark. The production of hadrons in the CFR can be described in a factorized framework by the convolution of Parton Distribution Functions (PDFs) and Fragmentation Functions (FFs)~\cite{Bacchetta:2006tn}. Here the PDFs describe the probability of finding a specific quark or gluon in a particular state inside the nucleon~\cite{Gao:2017yyd,Ethier:2020way} and the FFs dictate the formation of hadrons out of quarks and gluons~\cite{Metz:2016swz}. However, hadrons produced in the target-fragmentation region (TFR), formed under the participation of the spectator partons, are not described by this picture and have been largely unexplored until now.

This letter describes the first ever SIDIS detection of a hadron in the CFR (a $\pi^+$) in coincidence with a hadron in the TFR (a proton). The corresponding theoretical basis to study the TFR is based on the fracture function formalism and was established in Ref.~\cite{Trentadue:1993ka}
for the collinear case. This approach has been generalized to the spin- and transverse-momentum-dependent (STMD) case~\cite{Anselmino:2011ss}. Similar to the case of PDFs and FFs in the CFR, the fracture functions describe the conditional probability for the target remnant to form a specific final state hadron after the ejection of a particular quark. In electroproduction, the polarization state of the virtual photon depends on the longitudinal polarization of the lepton beam, which in turn selects preferentially one polarization state of the struck quark. The opposite polarization and transverse momentum of the remnant can introduce correlations between final-state hadrons produced in the TFR and hadrons produced in the CFR. The study of this dihadron production in SIDIS with a longitudinally polarized electron, where one hadron is produced in the CFR and another in the TFR, provides access to leading twist fracture functions~\cite{Anselmino:2011bb}.  
In the valence-quark region, the
polarization transfer from the beam to the active quark is expected to be significant at the relatively low center-of-mass energies accessible at CLAS12~\cite{Burkert:2020akg}. Preliminary studies using CLAS data indicated that these target-current correlations may be significant~\cite{Avakian:2016zos}. The high luminosity and high polarization of the electron beam along with a wide acceptance for the detection of many final-state particles makes CLAS12 an ideal place for studies of correlations between the target- and current-fragmentation regions.

Sizable beam single-spin asymmetries (SSAs) for a longitudinally polarized electron beam on nucleon targets and one or two hadrons detected in the CFR have been observed at JLab 
\cite{CLAS:2003qum,CLAS:2021opg,CLAS:2020igs,Hayward:2021psm},
HERMES~\cite{Airapetian:2006rx} and COMPASS~\cite{COMPASS:2014kcy,Parsamyan:2018ovx,Parsamyan:2018evv}. These results have been interpreted in terms of higher twist contributions, related to quark-gluon correlations. Here, higher twist refers to quantities that are suppressed by the hard scale of the process~\cite{Jaffe:1997vlv}.
When one of two hadrons is detected in the TFR and the other in the CFR, the beam SSAs measured here appear at leading twist~\cite{Anselmino:2012zz} without this suppression.

\begin{figure}
    \centering
    \includegraphics[width=0.45\textwidth]{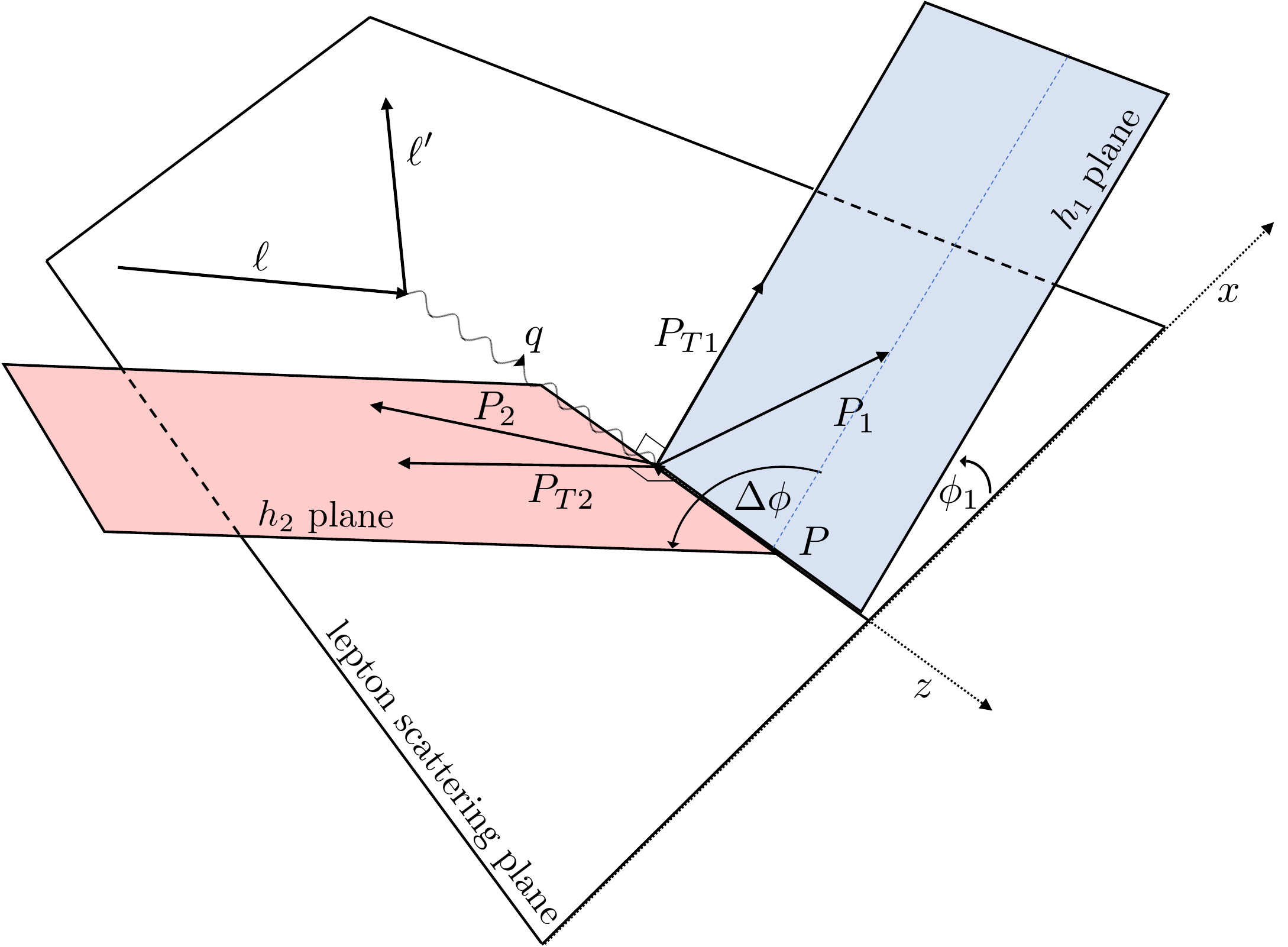}
    \caption{The SIDIS kinematics of back-to-back dihadron production in the center-of-mass frame. The $x$-$z$ plane is defined by the incoming and outgoing lepton with positive $z$ in the direction of the virtual photon. $\phi_1$ and $\phi_2$ are defined from the scattering plane to $P_1$ and $P_2$ in an anti-clockwise direction.}
    \label{fig:b2b_plane}
\end{figure}

In the target fragmentation region it is not possible to separate quark emission from hadron production. This prevents access to a chiral-odd quantity, such as the Collins function, to pair with any chiral-odd fracture functions in single hadron production and ultimately makes any chiral-odd fracture functions inaccessible in single hadron production. In contrast, in the double hadron production process $l(\ell) + N(P) \to l(\ell') + h_1(P_1) + h_2(P_2) + X$, at perturbative QCD leading order, the cross section expression includes all twist-2 fracture functions and quark fragmentation functions~\cite{Anselmino:2011ss,Anselmino:2011bb} by pairing the fracture function with a fragmentation function that dictates the production of a hadron in the CFR.

The process considered here is shown in Fig.~\ref{fig:b2b_plane}. We use the standard DIS variables: the momentum of the exchanged virtual photon, $q=l-l^\prime$, the scale of the process, $Q^2=-q^2$, the fractional longitudinal target momentum carried by the struck quark, $x=Q^2/2P \cdot q$, the fractional energy loss of the scattered electron $y=P \cdot q/P \cdot l$ and the hadronic mass of the system, $W^2=(P+q)^2$. The hadronic variables are defined below. For the case of a longitudinally polarized beam and unpolarized target after the integration over $\phi_2$ (the azimuthal angle of the TFR hadron) and keeping $\Delta\phi = \phi_2 - \phi_1$ fixed, there are two contributions to the cross section, $\sigma_{UU}$ and $\sigma_{LU}$ ~\cite{Anselmino:2011vkz},
\begin{align}
\sigma_{UU}  &= F_0^{{\hat u_1} \cdot D_1},
\label{Sig_UU}
\end{align}
\begin{align}
\sigma_{LU} & = \frac{ P_{{T1}} P_{{T2}}}{m_N m_2}  F_{{k1}}^{{\hat l_1}^{\perp h}\cdot D_1} \, \sin(\Delta\phi),
\label{Sig_LU}
\end{align}
where the structure functions $F_0^{\hat{u}_1 \cdot D_1}$ and $F_{k1}^{\hat{l}_1^{\perp h} \cdot D_1}$ are convolutions~\cite{Anselmino:2011bb} of the leading twist fracture functions $\hat{u}_1$ and $\hat{l}_1^{\perp h}$ with the unpolarized fragmentation function, $D_1$, which depend on the kinematic variables $x, Q^2, z_1, \zeta_2, P_{T1}^2,  P_{T2}^2$, and ${\Vec P}_{T1} \cdot {\Vec P}_{T2}$. The masses of the nucleon target, forward and backward produced hadrons are denoted as $m_N$, $m_1$ and $m_2$. The hadron 1, $h_1$, produced in the CFR ($x_{F1}>0$)
\footnote{We use the standard definition for Feynman  variables $x_{F1}$ and $x_{F2}$, see next section.}
is described by the standard scaled variable $z_1 = P{\cdot}P_1/P{\cdot}q$, describing the fraction of the virtual photon energy carried by the CFR hadron,  and its transverse momentum ${\Vec P}_{T1}$ (defined relative to the $q$ vector in the target rest frame) with magnitude $P_{T1}$ and azimuthal angle
$\phi_1$. The hadron 2, $h_2$, produced in the TFR ($x_{F2}<0$) is described by similar variables: the fractional longitudinal target momentum carried by the TFR hadron, $\zeta_2 \simeq E_2/E$ where $E$ is the energy of the target, 
and ${\Vec P}_{T2}$ ($P_{T2}$ and $\phi_2$) in the virtual photon and target nucleon center-of-mass frame. The usual hadronic scaling variable, $z$, is not used in the TFR because of ambiguities between soft hadron emission and target fragmentation~\cite{Anselmino:2011ss}.

In particular, the structure function in Eq.~\ref{Sig_LU} contains the fracture function $\hat{l}_{1}^{\perp h}(\zeta_2,P_{T2})$, describing the production of $h_2$ after the emission of a longitudinally polarized quark in an unpolarized nucleon and $D_1 (z_1, P_{T1})$, the unpolarized fragmentation function describing the formation of $h_1$. This structure function depends on the relative azimuthal angle of the two hadrons, indicating a long-range correlation between hadrons produced in the CFR and the TFR. The resulting beam-spin asymmetry contains the convolution of the fracture function and the fragmentation function modulated by $\sin\Delta\phi$,
\begin{align}
      {\mathcal A}_{LU} &= - \sqrt{1-\epsilon^2} \frac{\vert \Vec P_{T1} \vert \vert 
\Vec P_{T2} \vert}{m_N \, m_2} 
\, \frac{{\mathcal C} [w_5 \, \hat{l}_1^{\perp h} D_1]}
{{\mathcal C} [\hat{u}_1 D_1]} 
\sin \Delta \phi.
\label{ptdep}
\end{align}
The depolarization factor in the front of Eq.~\ref{ptdep}, governing the polarization transfer from the electron to virtual photon, is described by the variable
\begin{equation}
\epsilon = \frac{1-y-\frac{1}{4}\gamma^2 y^2}{1-y+\frac{1}{2}y^2+\frac{1}{4}\gamma^2y^2},    
\end{equation}
with $\gamma = 2m_N x/Q$. The weight factor is given by
\begin{align}
  w_{5} &=  \frac{(\Vec k_{\perp} \cdot \Vec P_{T2}) (\Vec P_{T1} 
\cdot \Vec P_{T2}) - (\Vec k_{\perp} \cdot \Vec P_{T1}) 
\Vec P_{T2}^2}{(\Vec P_{T1} \cdot \Vec P_{T2})^2
- \Vec P_{T1}^2 \Vec P_{T2}^2}, \>
\end{align}
and the following notation is used for the transverse momentum convolution 
\begin{align}
    {\mathcal C} \, [f (\Vec k_{\perp}, &\Vec k'_{\perp}, \ldots)] = \sum_a e_a^2 \, x_B \,  \int d^2 \Vec k_{\perp} \, \int d^2 \Vec k_{\perp}' \times 
\nonumber \\
& \delta^2 (\Vec k_{\perp} - \Vec k_{\perp}' - \Vec P_{T1}/z_1) 
 \, f (\Vec k_{\perp}, \Vec k'_{\perp}, \ldots) \>,
\end{align}
where $k_\perp$ is the transverse momentum of the initial quark with respect to the virtual photon, $k_\perp'$ is the transverse momentum after the interaction and the summation runs over the quark flavors. 

\section{Experiment}
The data, corresponding to SIDIS events with a $\pi^+$ in the CFR and proton in the TFR, were taken in two run periods in the fall of 2018 and spring 2019 using 10.6 and 10.2 GeV longitudinally polarized electron beams delivered by the Continuous Electron Beam Accelerator Facility at Jefferson Lab~\cite{Leemann:2001dg}. The electron beam was incident on a liquid-hydrogen target and reactions were recorded using the CLAS12 spectrometer~\cite{Burkert:2020akg}. The beam polarization averaged to $85.7\pm 1.6\%$ and was flipped at a rate of 30~Hz to minimize systematic effects. 

A tracking subsystem consisting of drift chambers in a toroidal magnetic field was used to identify and reconstruct particles scattered in the forward direction. A high-threshold Cherenkov counter was used to distinguish between electrons and final-state hadrons. Additional identification criteria for the electrons was also imposed using a series of electromagnetic calorimeters. The CLAS12 forward time-of-flight systems, composed of six arrays of plastic scintillation counters, were used to analyze the velocity vs. momentum relationship of positive tracks to distinguish between hadron species. The pions were limited to momenta $1.2~<~p~<~4.0$~GeV in order to avoid regions of low efficiency (lower limit) and minimize misidentification of kaons (upper limit). Protons were required to have a momentum greater than 0.5~GeV (no strict upper limit was enforced but the distribution of upper proton momenta dies down around 2.5~GeV, well before any significant contamination from lighter hadron species). The reconstructed electron and hadrons were required to have been identified in the so-called ``forward detector'' of CLAS12 and a requirement has been placed on the polar angle of each track, $\theta < 30^\circ$.

SIDIS events were selected with the usual requirement that $Q^2>1$~GeV$^2$ and the mass of the hadronic final-state, $W>2$ GeV. Events with a radiated photon were limited by requiring events to have $y~<~0.75$. At energies accessible by fixed target experiments there is no rapidity gap and the forward and backward regions were defined by the variable $x_F$, in the virtual photon-nucleon center-of-mass frame, with the requirement that ${x_F}_1~>~0$ and ${x_F}_2~<~0$. The Feynman-$x$ variable is defined as  ${x_F}_{1(2)}=2{P_\parallel}_{1(2)} / W$, where $P_\parallel$ is the longitudinal momentum of the hadron and takes a positive value if the hadron moves in the same direction as the momentum transfer dictated by virtual photon and a negative value if it moves in the same direction of the target in the center-of-mass frame.
An additional requirement on the boost-invariant quantity $\Delta Y \equiv Y_1 - Y_2 > 0$, where $Y$ is the rapidity evaluated in the Breit frame defined as $2Y_{1(2)}=\ln(E_{1(2)}+P_{\parallel1(2)})/(E_{1(2)}-P_{\parallel1(2)})$, was required to enforce separation between the hadrons. The asymmetries were studied as a function of both of these variables to investigate the transition from one region to another. Contributions from $\Delta^{++}$ decays were minimized by requiring the invariant mass of the observed hadrons to have $M_{p\pi} > 1.5$~GeV.
Finally, the missing mass of the process $ep~\rightarrow~e'p \pi^+X$ was restricted to be greater than 0.95~GeV in order to avoid contributions from diffractive meson production.

\section{Results}
\begin{figure}
    \centering
    \includegraphics[width=0.48\textwidth]{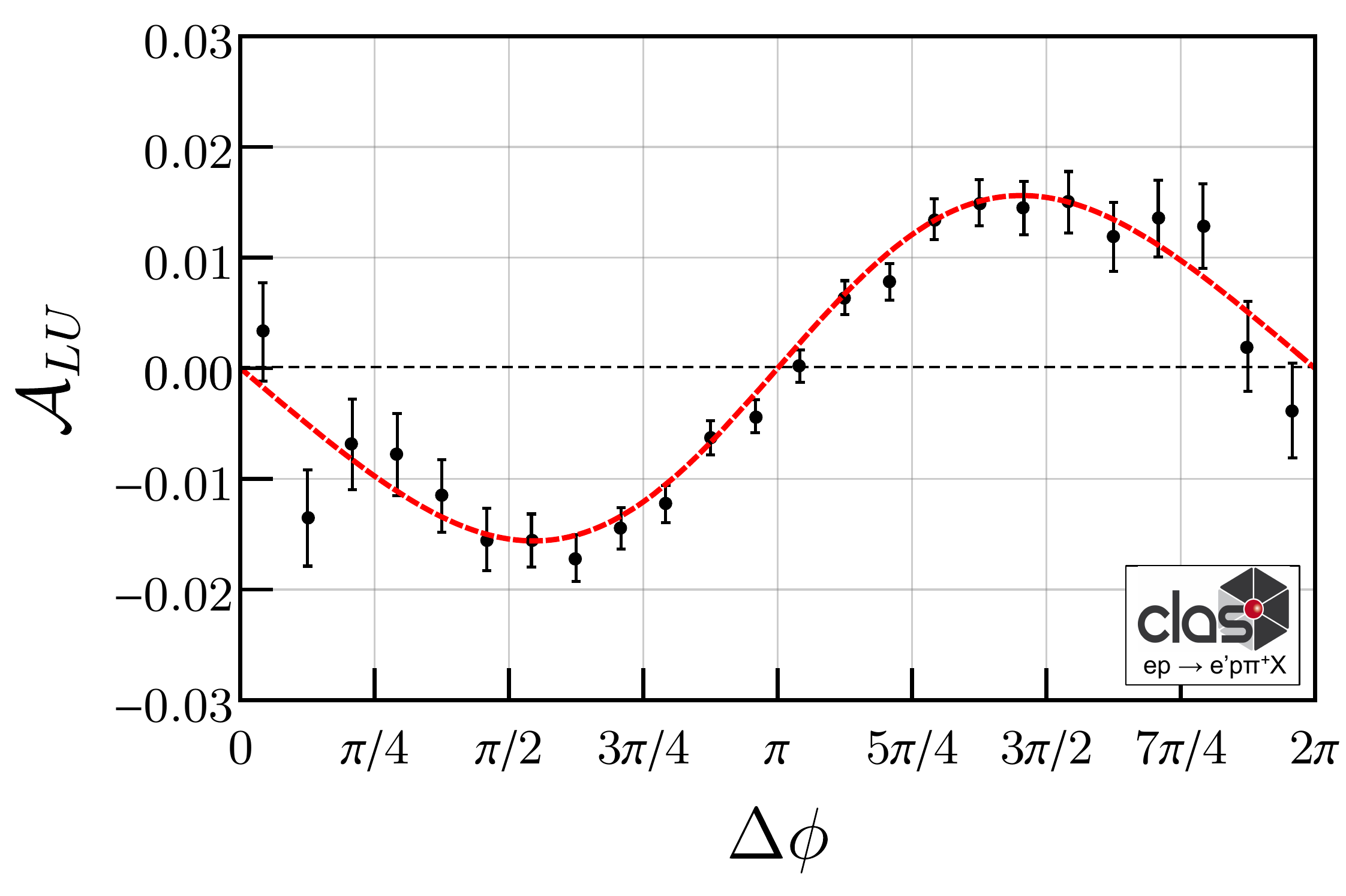}
    \caption{The beam spin asymmetry, $\mathcal{A}_{LU}$, as a function of $\Delta \phi$ and integrated over all other kinematics for the entire data set. A clear $\sin(\Delta\phi)$ dependence is observed with small $\sin(2\Delta\phi)$ contributions.}
    \label{fig:sinusoid}
\end{figure}

The beam-spin asymmetry can be accessed using the yields, $N^\pm$, of events with a proton in the backward region and a positive pion in the forward region, produced from the scattering of an electron with helicity $\pm$, written
\begin{align}
\label{eq:fit}
  & \mathcal{A}_{LU}(\Delta\phi)=
  \frac{1}{P_\text{beam}} \frac{N^+(\Delta\phi)- N^-(\Delta\phi)}{N^+(\Delta\phi)+ N^-(\Delta\phi)}=\\
     & \mathcal{A}_{LU}^{\sin\left(\Delta\phi\right)}\sin(\Delta\phi)+\mathcal{A}_{LU}^{\sin\left(2\Delta\phi\right)}\sin(2\Delta\phi)\nonumber  \nonumber,
\end{align}
with the dependence on $\sin(\Delta\phi)$ and $\sin(2\Delta\phi)$ described in Ref.~~\cite{Anselmino:2012zz}, and fitting for the resulting azimuthal modulation amplitudes. The beam polarization, $P_\text{beam}$, has been divided out of the asymmetries. The amplitudes in Eq.~\eqref{eq:fit} were extracted from the data using an unbinned maximum likelihood fit that includes both modulations of $\mathcal{A}_{LU}$. 
A binned $\chi^2$-minimization fit with $9$ bins in $\Delta\phi$ was also performed and is in very good agreement with the unbinned fit with a mean reduced $\chi^2$ of 1.01. The count-rate asymmetry between positive and negative electron helicities as a function of $\Delta\phi$ is shown in Fig.~\ref{fig:sinusoid}. The count rate difference exhibits a clear $\sin(\Delta\phi)$ behavior with a much smaller $\sin(2\Delta\phi)$ contribution.

\begin{figure}
    \centering
    \includegraphics[width=0.48\textwidth]{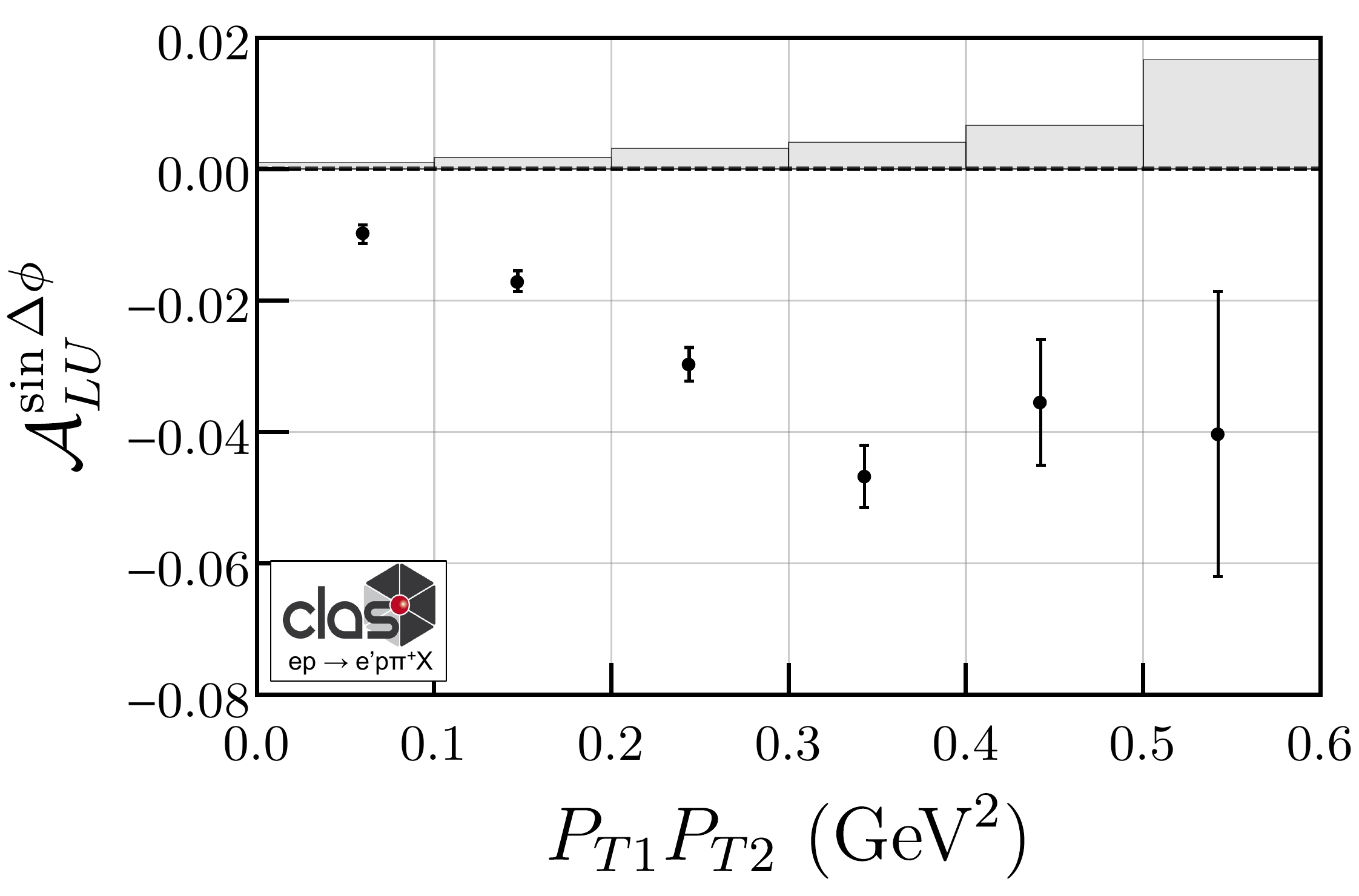}
    \caption{The measured $\mathcal{A}_{LU}^{\sin\Delta\phi}$ asymmetry as a function of $P_{T1} P_{T2}$. Thin black bars indicate statistical uncertainties and wide gray bars represent systematic uncertainties.}
    \label{fig:pTpT}
\end{figure}

The dependence of $\mathcal{A}_{LU}^{\sin\Delta\phi}$ on the product of transverse momenta of the proton and pion is shown in Fig.~\ref{fig:pTpT} and is consistent with a linear increase in magnitude and approaches zero as the transverse momentum goes to zero, following the kinematic dependence predicted by theory (see. Eq.~\ref{ptdep}). Additional multidimensional asymmetries as a function of the product of the transverse momenta in bins of $z_1$ are given in the supplementary material.

Due to the correlation between different kinematic variables and the product of the transverse momenta of both hadrons, the asymmetries can be weighted by dividing out the depolarization and kinematic weighting factor in Eq.~\ref{ptdep}, $\sqrt{1-\epsilon^2}(P_{T1} P_{T2}) / (m_N m_2)$, (note that for this measurement $m_N = m_2 = 0.938$~GeV). The dependence of the resulting ratio, which should directly depend on the ratio of the convolutions of fracture and fragmentation functions, was studied for several different kinematic variables. 

\begin{figure}
    \centering
    \includegraphics[width=0.48\textwidth]{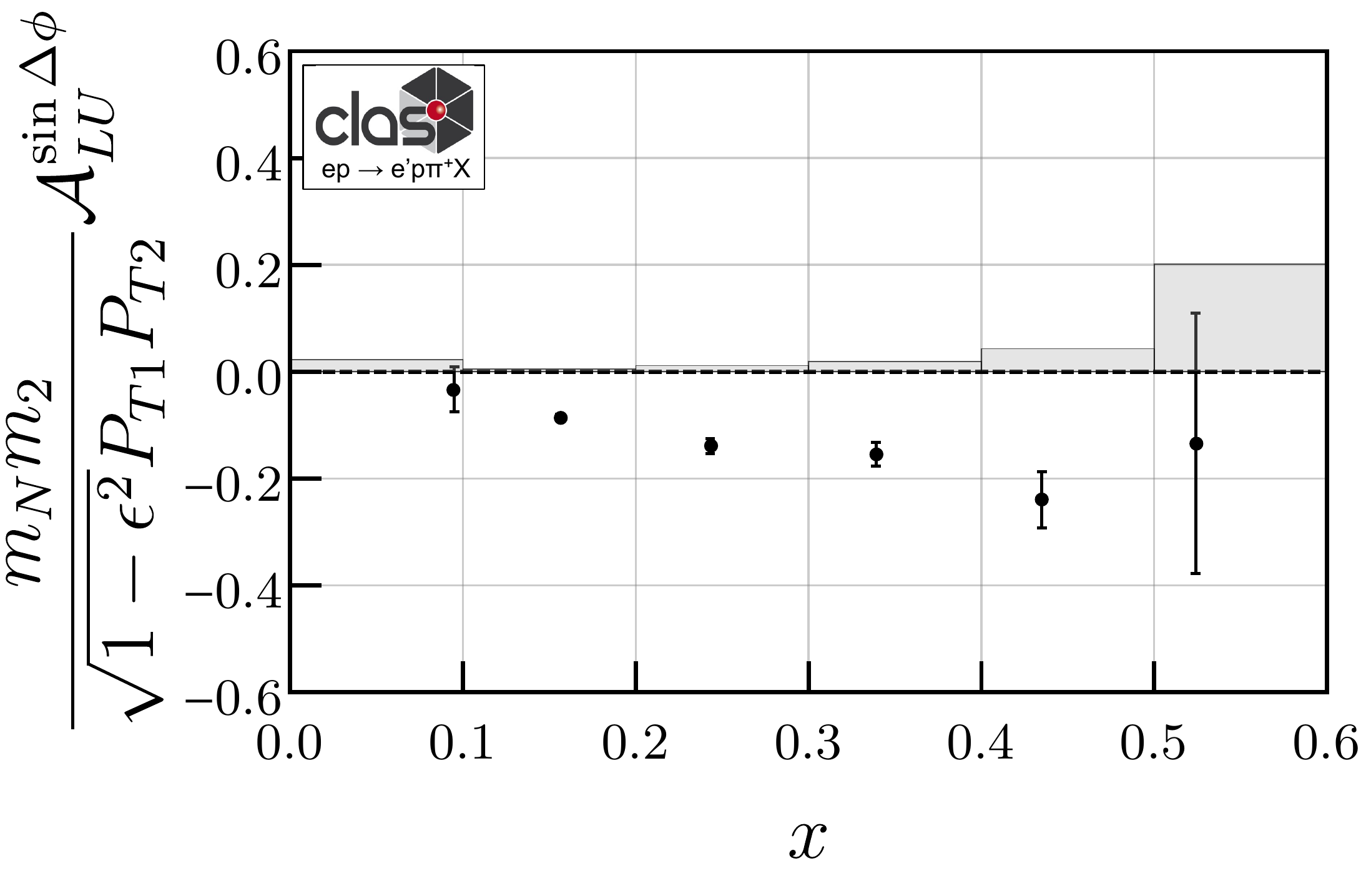}
    \caption{The measured weighted $\mathcal{A}_{{LU}}^{\sin\Delta\phi}$ asymmetry as a function of $x$. Thin black bars indicate statistical uncertainties and wide gray bars represent systematic uncertainties.}
    \label{fig:x}
\end{figure}

\begin{figure}
    \centering
    \includegraphics[width=0.48\textwidth]{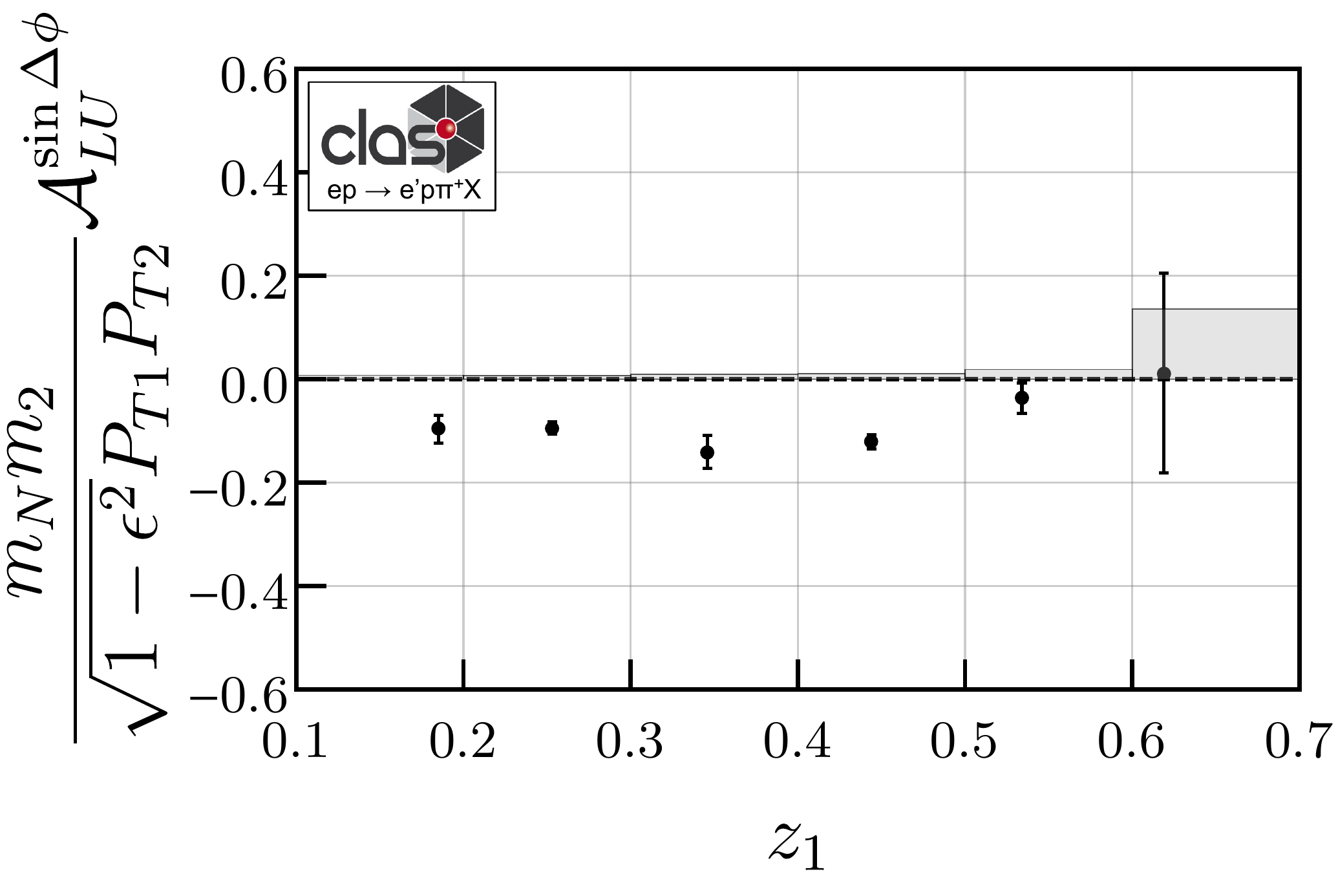}
    \caption{The measured weighted $\mathcal{A}_{{LU}}^{\sin\Delta\phi}$ asymmetry as a function of $z_1$. Thin black bars indicate statistical uncertainties and wide gray bars represent systematic uncertainties.}
    \label{fig:z_pi}
\end{figure}

\begin{figure}
    \centering
    \includegraphics[width=0.48\textwidth]{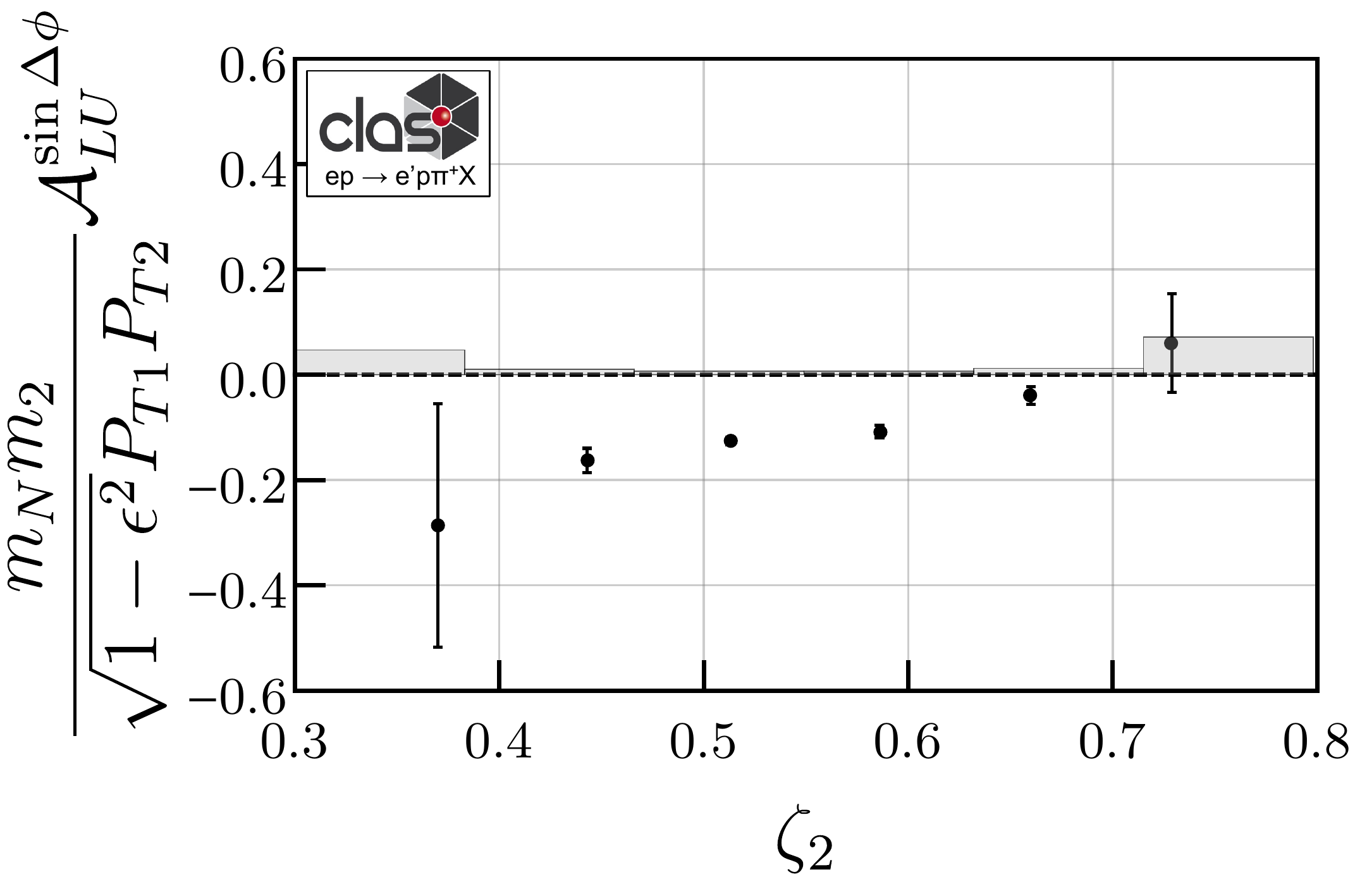}
    \caption{The measured weighted $\mathcal{A}_{{LU}}^{\sin\Delta\phi}$ asymmetry as a function of $\zeta_2$. Thin black bars indicate statistical uncertainties and wide gray bars represent systematic uncertainties.}
    \label{fig:zeta}
\end{figure}

The $x$-dependence, shown in Fig.~\ref{fig:x}, has the general trend of increasing in magnitude as $x$ increases. This strong dependence implies that the correlation of final-states hadrons is most significant in the valence quark region. The dependence on $z_{1}$ of the pion, which reflects the fragmentation function dependence, is shown in Fig.~\ref{fig:z_pi}. At relatively small $z_1$, contributions from the initial quark transverse momentum can be neglected and the main contribution to the produced hadron transverse momentum comes from the struck quark hadronization process. Indeed, this dependence appears relatively flat, with a possible decrease at higher values of $z_1$ where effects from decreasing transverse momentum begin to dominate. This relatively weak dependence may also be a consequence of cancellation between the pion fragmentation functions in the numerator and denominator. The dependence on $\zeta_2$, shown in Fig.~\ref{fig:zeta}, is stronger and may be interpreted in terms of strong correlations with other variables such as $x$; typically the higher longitudinal momentum carried by the struck quark, the lower the longitudinal momentum available for the TFR hadron. 

Additional kinematic dependences are included in the supplementary material. The asymmetries plotted versus ${x_F}_1$ and $\Delta Y$ (with the kinematic constraints on ${x_F}_2$, ${x_F}_1$ and $\Delta Y$ removed) show a relatively flat dependence in the ${x_F}_1 < 0$ region with a possible transition in the positive region where the proton is increasingly likely to have originated in the CFR and the back-to-back fracture function formalism no longer holds. Finally, the dependence on the missing mass, with the $M_x > 0.9$~GeV requirement removed, shows a relatively flat behavior above the contributions from diffractive $ep \rightarrow e'p\pi^+\pi^-$ and $ep \rightarrow e'p\pi^+\rho^-$ events.

Relevant systematic uncertainties have been estimated using a number of methods. Monte Carlo simulations were performed using the PEPSI generator~\cite{Mankiewicz:1991dp} and a GEANT4-based simulation~\cite{Agostinelli:2002hh,Ungaro:2020xlc} of the detector for acceptance, efficiency and particle-ID studies. Good agreement for all underlying variables was observed. Systematic uncertainties due to bin migration and the scale uncertainty on the beam polarization can reach a few percent in each bin. Other effects due to particle misidentification, accidental coincidences, photoproduction of electrons and contamination from target-fragmentation pions (baryonic decays like $\Delta^{++} \rightarrow p\pi^+$) have all been estimated to be small. Contributions from radiative effects are avoided by limiting our kinematic range and are estimated to be small. However, more theory effort will be needed in future for detailed studies of radiative corrections in SIDIS in general, and back-to-back dihadron production in particular.

The largest systematic uncertainty comes from contributions from the unpolarized cross section. After integration over $\phi_2$, the only remaining structure functions are $F_{UU}$ and $F^{\sin(\Delta\phi)}_{LU}$. However, due to the non-perfect acceptance of CLAS12 and the potential resulting non-orthogonality of modulations, the other unpolarized structure functions may impact our extraction of the $F_{LU}$ amplitudes. Since there is no experimental data on the spin-independent ($UU$) modulations, the uncertainty due to not including these modulations in our fit has been evaluated using Monte Carlo simulations by injecting various values of the amplitudes and then performing fits not including the $UU$ modulations in order to measure the possible deviation in our $LU$ amplitudes. The effect is heavily bin-dependent, but can only reach magnitudes similar to the statistical uncertainty at edges of kinematic space. For example, in the lowest $x$-bin we assign an uncertainty to $A_{LU}^{\sin\Delta\phi}$ of 0.006 compared to 0.021 at the highest $x$ bin where the measurement is less constrained. This study is both dependent on our injected amplitudes and fairly conservative and so, therefore, may correspond to an overestimate.

\section{Conclusions}
In summary, the kinematic dependences of beam SSAs in the production of two hadrons in opposite hemispheres have been measured for the first time. The asymmetries may be interpreted in a framework described by TMD factorization into fracture functions and fragmentation functions, with the  conditional probability of finding a proton originating from the target remnant after the emission of a quark which undergoes hadronization to form a final-state $\pi^+$. The $P_T$ dependence of the asymmetries is consistent with predictions of the factorized framework and can ultimately be used to test TDM factorization once extractions of the relevant functions are available and predictions have been made.

Our measurement of correlations between the target- and current-fragmentation regions develop a new methodology to quantify the relationship between the spin and transverse momenta of quarks in the nucleon and provides a new avenue for studies of the complex nucleonic structure in terms of quark and gluon degrees of freedom. The kinematics of the generated asymmetries are not in the perturbative regime, but instead, the asymmetries likely originate from correlations between the longitudinally polarized struck quark's azimuthal angle and the azimuthal angle of the proton produced in the TFR~\cite{Anselmino:2011ss}.

Future work will extend the analysis to other hadron species in both the TFR and CFR in order to test the universal nature of fracture functions. The flavor dependence of fracture functions can be extracted by comparing with deuterium targets, of which comparable statistics to the proton-target data shown here have already been collected by CLAS12. Finally, polarized-target measurements will enable access to the complete set of leading-twist fracture functions, which cannot be observed in single-hadron production but require an additional hadron to produced in the CFR like the measurement performed here.

\section{Acknowledgements}
We acknowledge the outstanding efforts of the staff of the Accelerator and the Physics Divisions at Jefferson Lab in making this experiment possible. This work was supported in part by the U.S. Department of Energy, the National Science Foundation (NSF), the Italian Istituto Nazionale di Fisica Nucleare (INFN), the French Centre National de la Recherche Scientifique (CNRS), the French Commissariat pour l$^{\prime}$Energie Atomique, the UK Science and Technology Facilities Council, the National Research Foundation (NRF) of Korea, the Helmholtz-Forschungsakademie Hessen für FAIR (HFHF), the Ministry of Science and Higher Education of the Russian Federation, the Chilean Agencia Nacional de Investigacion y Desarrollo (ANID) via grant PIA/APOYO AFB180002 and the European Union's
Horizon 2020 Research and Innovation program under Grant
Agreement N.824093 (STRONG2020). The Southeastern Universities Research Association (SURA) operates the Thomas Jefferson National Accelerator Facility for the U.S. Department of Energy under Contract No. DE-AC05-06OR23177. The work is also supported in part by DOE grant DE-FG02-04ER41309.

\bibliographystyle{apsrev4-1}
\bibliography{references}
\end{document}